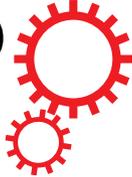

# Interfacial undercooling in solidification of colloidal suspensions: analyses with quantitative measurements

Jiaxue You[1], Lilin Wang[2], Zhijun Wang[1], Junjie Li[1], Jincheng Wang[1], Xin Lin[1] & Weidong Huang[1]

Interfacial undercooling in the complex solidification of colloidal suspensions is of significance and remains a puzzling problem. Two types of interfacial undercooling are supposed to be involved in the freezing of colloidal suspensions, i.e., solute constitutional supercooling (SCS) caused by additives in the solvent and particulate constitutional supercooling (PCS) caused by particles. However, quantitative identification of the interfacial undercooling in the solidification of colloidal suspensions, is still absent; thus, the question of which type of undercooling is dominant in this complex system remains unanswered. Here, we quantitatively measured the static and dynamic interface undercoolings of SCS and PCS in ideal and practical colloidal systems. We show that the interfacial undercooling primarily comes from SCS caused by the additives in the solvent, while PCS is minor. This finding implies that the thermodynamic effect of particles from the PCS is not the fundamental physical mechanism for pattern formation of cellular growth and lamellar structure in the solidification of colloidal suspensions, a general case of ice-templating method. Instead, the patterns in the ice-templating method can be controlled effectively by adjusting the additives.

The solidification of colloidal suspensions is commonly encountered in a variety of natural processes such as the growth of sea ice[1] and frost heave[2], and engineering situations such as cryobiology[3], tissue engineering[4], ice-templating bio-inspired porous materials and composites[5–19], thermal energy storage[20] and soil remediation[21]. In particular, ice-templating porous materials have attracted increasing attention due to the novel micro-aligned structures that can be easily produced for a wide range of applications[5–16,18].

One of the key issues therein is the pattern formation. The formation of microstructures is closely related to the interfacial instability during the freezing of colloidal suspensions and the subsequent development of interfacial morphologies[22,23]. The interfacial instability during freezing strongly depends on the interfacial undercooling, which has been extensively revealed by the research community in studies of solidification[24,25]. Accordingly, it is believed that the interfacial undercooling is also of significance in microscopic pattern formation of freezing colloidal suspensions.

Two types of interfacial undercooling have been proposed to occur in the solidification of colloidal suspensions, i.e., solute constitutional supercooling (SCS) caused by additives in the solvent[12,13,17], and particulate constitutional supercooling (PCS) caused by particles[22,23]. The theory of SCS is based on the classical alloy solidification principle[12,24,25], while the theory of PCS is derived from multi-particle thermodynamics[22], a typical characteristic of the colloidal suspensions system. In the past decade, since its initial proposal, the PCS theory has attracted the extensive attention of researchers in many fields, such as porous ceramics[26], polymers and composites[27,28], bone tissue engineering[29], metal–ceramic composites[30], the science of soft matter[31,32], geophysical science[33], thermal energy storage[34], crystal growth[35], cryobiology[36], etc. To date, there has been no report on the quantitative measurements of interfacial undercooling during the solidification of colloidal suspensions, much less on the distinctions between these two types of interfacial undercooling.

[1]State Key Laboratory of Solidification Processing, Northwestern Polytechnical University, Xi'an 710072, P.R. China. [2]School of Materials Science and Engineering, Xi'an University of Technology, Xi'an 710048, P.R. China. Correspondence and requests for materials should be addressed to Z.W. (email: zhijwang@nwpu.edu.cn) or J.W. (email: jchwang@nwpu.edu.cn)

The question of which type of undercooling is dominant in the solidification of colloidal suspensions remains an unsolved but important issue. First, determining the individual effects of SCS and PCS can improve the fundamental understanding of the physical mechanism of pattern formation during the freezing of colloidal suspensions. Moreover, this determining the individual effects will further pave the way toward controlling the microscopic pattern formation in this complex system. For example, if the PCS is dominant, then the freezing pattern can be modulated *via* the particle size or particle shape; otherwise, it can be adjusted by changing the additives. Here, we quantitatively measured the static and dynamic interfacial undercoolings of SCS and PCS in ideal and practical colloidal systems.

In this letter, interfacial undercoolings were quantitatively measured based on a novel experimental method. The contributions of SCS and PCS were confirmed for the first time. A detailed description of the experimental apparatus and gauging method is given in ref. 37. A sketch of this method is shown in Fig. S1 (Supplementary Information). In the method, the interfacial undercooling is visualized through the discrepancy of solid/liquid interfacial positions in two adjacent Hele-Shaw cells of the colloidal suspension and its compared counterpart in a uniform thermal gradient apparatus. The thermal gradient plays a key role on measuring interfacial undercoolings and positions. In order to reduce radiative and convective perturbations, the thermal blocks were covered with heat-insulating shield and the gap between hot and cold copper blocks was covered with double-glazed windows, which had both excellent thermal insulation and optical microscope observation. More importantly, these two Hele-Shaw cells (the composition of their walls is glass, with a cross-section of 2 mm × 0.1 mm) were placed tightly on a sufficiently large plane glass plate (with a cross-section of 30 mm × 0.2 mm) in order to obtain a fixed linear thermal gradient. The thermal gradient in Hele-Shaw cells is determined by heat conduction of the plane glass plate so as to minimize the effect of the difference in the thermal conductivities of particles, liquid water and ice on the measurements of interfacial undercoolings. The validity of the measurements is verified based on tests with different thermal gradients. The SCS and PCS can be well distinguished by designing different compared counterparts. After quantitatively measuring the SCS and PCS in different systems of colloidal suspensions, we analyzed the results based on theoretical predictions. In the quantitative measurements, ideal systems of polystyrene microsphere (PS) suspensions were first considered. After discovering the minor contribution of PCS, we also measured the interfacial undercooling in practical systems of α-alumina suspensions with both static and dynamic interface to further confirm the contributions of SCS and PCS.

The first system we chose is PS suspensions (Bangs Lab, USA). The nominal solvent of PS suspensions is deionized water. The density of PS particles is almost the same as that of water. The mean diameter of the particles is $d = 1.73\,\mu m$ with a poly-dispersity smaller than 5%, and the initial volume fraction of particles is $\phi_0 = 33\%$. The PS suspensions system is stable only with weak sedimentation, i.e., it is an ideal system to investigate the freezing of colloidal suspensions. Although the solvent of deionized water is marked on the nominal label of PS suspensions, we believe that there are still very small quantities of residual solutes from the synthesis of PS particles, even after great efforts of purification in these commercial PS suspensions. The residual solutes will also cause SCS during the freezing of PS suspensions. Therefore, in the measurement, first, we verified this type of SCS by comparing deionized water with the supernatant from the PS suspensions by centrifugation. Furthermore, we compared each PS suspension with its supernatant to confirm the individual contribution of PCS. The combination of SCS and PCS accounts for the whole interfacial undercooling during the solidification of colloidal suspensions.

Figure 1(a) shows the measurement of SCS through the interface position comparison between the deionized water (left cell of Fig. 1(a)) and the supernatant (right cell of Fig. 1(a)) within a microscopic image. The upper end of the cell is the heating zone, while the lower end of the cell is the cooling zone, which builds a linear thermal gradient $G = 7.23\,K/cm$. The pulling speed V is 0. In Fig. 1(a), the position of solid/liquid interface in the cell of deionized water is much higher than that of the supernatant, which indicates that the freezing point of the deionized water is much higher than that of the supernatant. The discrepancy of the solid/liquid interface positions between the deionized water and the supernatant is 171 μm, corresponding a SCS of 0.123 K with $G = 7.23\,K/cm$.

Comparison of the interfacial position between the colloidal suspension and its supernatant is shown in Fig. 1(b), which exhibits the measurement of PCS. The interfacial position of the supernatant is almost identical to that of its suspension, which means that the freezing point of the supernatant is almost the same as that of its suspensions. Therefore, the PCS is almost undetectable and should be smaller than 0.01 K if it exists in this PS colloidal suspensions system (the precision of the experimental method has been demonstrated to be 0.01 K[37]).

Consequently, in Fig. 1, the interfacial undercooling of colloidal suspensions mainly comes from SCS. To further confirm this conclusion, the interfacial undercoolings of PS suspensions systems with particles of different diameters and different volume fractions were measured. All the results are similar to that in Fig. 1. The comparisons of interface positions are shown in Fig. S2 (Supplementary Information), and the interfacial undercoolings of SCS and PCS are shown in Table 1. Surprisingly, all the results indicate that the PCS makes minor contribution to the interfacial undercooling of PS colloidal suspensions. However, a 5K PCS was reported in ref. 23 under $d = 1\,\mu m$ and $\phi_0 = 50\%$, which were similar to our test conditions, $d = 1.73\,\mu m$ and $\phi_0 = 33\%$. These two testing results (5K and less than 0.01 K) are obviously divergent.

These unexpected results deserve further analysis in considering the PCS theory[22,23,38,39]. Recently, PCS theory was first proposed and applied to address some puzzling phenomena with numerous unexplained features in freezing colloidal suspensions[40,41]. In PCS theory, the particle-controlled interfacial undercooling is described as

$$\Delta T_{PCS} = T_m - T_f = T_m \frac{\Pi(\phi)}{\rho_w L_f},\qquad(1)$$

where $\Pi(\phi) = \frac{\phi}{v_p} k_B T_m Z(\phi)$ is the osmotic pressure caused by the concentrated layer of particles, and $Z(\phi) = \frac{1 + a_1\phi + a_2\phi^2 + a_3\phi^3 + a_4\phi^4}{1 - \phi/\phi_p}$ is the dimensionless compressibility factor. $T_m$ is the melting point of ice. $T_f$ is the depressed melting point. $\rho_w$ is the density of water. $L_f$ is the latent heat, and $\phi$ is the volume fraction of particles. $k_B$ is the Boltzmann constant and $v_p = \frac{4\pi}{3}\left(\frac{d}{2}\right)^3$ is the volume of a particle. The maximum volume fraction of particles is $\phi_p = 0.64$, $0 \leq \phi \leq 0.64$. $a_1$, $a_2$, $a_3$ and $a_4$ are fitting parameters of $\Pi$. In this theory, the PCS comes from the depressed equilibrium melting point caused by the osmotic pressure of concentrated particles ahead of the freezing interface, which includes consolidated physical foundations[38]. However, the determination of the dimensionless compressibility factor in osmotic pressure is casual in refs 23 and 39. The variation of $Z(\phi)$ with different $\phi$ has been well investigated[42,43]. Originally, the fitting parameters from refs 22,38,43 and 44 are

$$a_1 = 2.4375, \quad a_2 = 3.75, \quad a_3 = 2.375, \quad \text{and } a_4 = -14.1552.$$

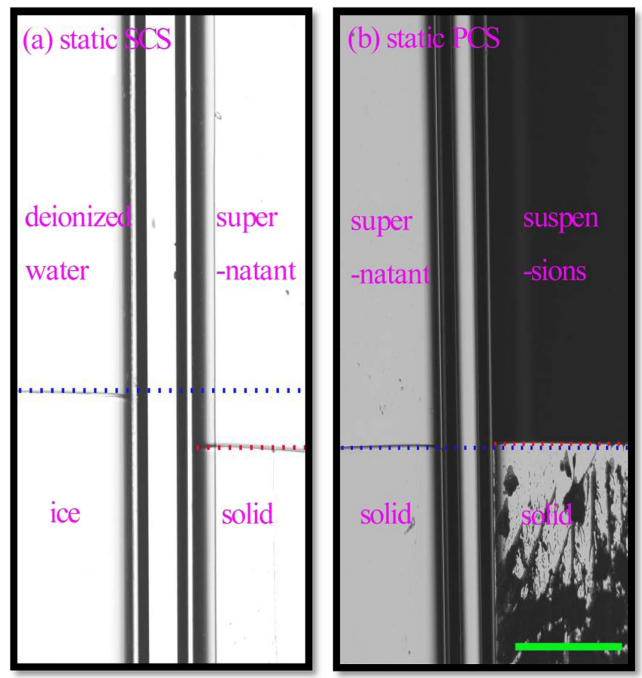

**Figure 1.** The static interfacial positions in two side-by-side Hele-Shaw cells of the deionized water and the supernatant from PS colloidal suspensions of d = 1.73 μm, $\phi_0$ = 33% (**a**); and two side-by-side Hele-Shaw cells of the colloidal suspensions and its supernatant (**b**) in a uniform thermal gradient of G = 7.23 K/cm. The distances between the static interfacial positions reveal the static interfacial undercoolings. The pulling speed is V = 0. The scale bar is 200 μm.

| | PS colloidal suspensions | | |
|---|---|---|---|
| d (μm) | 1 | 1.73 | |
| $\phi_0$ | 20% | 20% | 33% |
| Measured SCS ($10^{-2}$ K) | 12.1 ± 0.5 | 12.3 ± 0.4 | 12.3 ± 0.4 |
| Measured PCS ($10^{-2}$ K) | 0 ± 0.18 | 0 ± 1.87 | 0 ± 0.16 |
| PCS theoretical predictions ($10^{-2}$ K)   A | $3.06 \times 10^{-7}$ | $5.91 \times 10^{-8}$ | $1.69 \times 10^{-8}$ |
|   B | 17.2 | 3.34 | 19.5 |

**Table 1.  Static undercoolings from measurements and predictions.** Prediction A is from refs 22 and 38, while prediction B comes from refs 23 and 39. For PS colloidal suspensions of different d and $\phi_0$.

However, the fitting parameters used in refs 23 and 39 were

$$a_1 = 1 \times 10^7, \quad a_2 = 2 \times 10^9, \quad a_3 = 3 \times 10^9, \quad \text{and } a_4 = -8 \times 10^9,$$

in order to give way to the experimental results of filtration pressure of drilling fluid filtercakes[23]. Subsequently, experimental depressions of the freezing point were used to confirm these huge fitting parameters. However, the experimental data of depressed freezing points cannot be used to confirm the predicted PCS. First, the colloidal suspensions contained a large number of ions which can dramatically depress the freezing point of water[45]. Second, the bentonite used was a mixture of a variety of particles with different sizes, while $\Pi$ greatly depends on the particle radius, with an inverse proportion to the third power of the particle radius and thus PCS is inversely proportional to the third power of the particle radius.

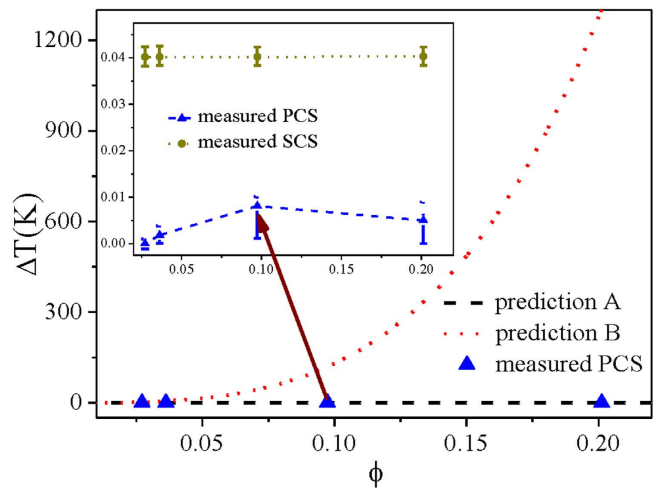

**Figure 2. Measured PCS compared with the theoretical PCS.** Prediction A of PCS is from refs 22 and 38, while prediction B of PCS comes from refs 23 and 39. The inset is the measured value of SCS and PCS for alumina suspensions with d = 50 nm under G = 7.23 K/cm and V = 0.

| | alumina suspensions | | | |
|---|---|---|---|---|
| d (μm) | 0.05 | | | |
| $\phi_0$ | 2.72% | 3.63% | 9.74% | 20.12% |
| Measured SCS ($10^{-2}$ K) | 4.01 ± 0.42 | 4.02 ± 0.42 | 4.03 ± 0.42 | 4.02 ± 0.42 |
| Measured PCS ($10^{-2}$ K) | 0 ± 0.2 | 0.18 ± 0.22 | 0.91 ± 0.85 | 0 ± 0.90 |
| PCS theoretical predictions ($10^{-2}$ K) A | $1.93 \times 10^{-4}$ | $6.98 \times 10^{-4}$ | $7.45 \times 10^{-4}$ | $2.24 \times 10^{-3}$ |
| B | 260 | 1640 | 12830 | 127701 |

**Table 2. Static undercoolings from measurements and predictions.** Prediction A is from refs 22 and 38, while prediction B comes from refs 23 and 39. For alumina suspensions of different $\phi_0$.

By using the original fitting parameter for Z(ϕ), the theoretical PCS in the PS systems investigated here is approximately $10^{-9}$ K, shown as prediction A in Table 1, which is too small to be detected. Both these theoretical results and the present measurements demonstrated that the PCS is minor compared with the SCS.

In PCS theory, the PCS is inversely proportional to the third power of the particle radius. Moreover, because the α−alumina suspensions are quite commonly used in the preparing of ice-templating porous ceramics, the interfacial undercoolings in the freezing of α−alumina colloidal suspensions were further measured. The α−alumina powder with a mean diameter d = 50 nm and a density of 3.97 g cm⁻³ are used (Wanjing New Material, Hangzhou, China, ≥99.95% purity, monodispersity). The alumina suspensions were prepared by using HCl (hydrogen chloride) and deionized water as the solvent following ref. 46. Initial volume fractions were $\phi_0$ = 2.72%, 3.63%, 9.74% and 20.12% (wt% = 10, 13, 30 and 50) in four different systems, respectively. These measurements were also made under the conditions of thermal gradient G = 7.23 K/cm and pulling speed V = 0. The SCS and PCS from the measurements are shown in Fig. 2. The interface position comparisons of the static SCS and PCS for alumina suspensions are shown in Fig. S3 (Supplementary Information). The measured PCS is still extremely small (blue triangular points in Fig. 2), i.e., it makes minor contribution to the total interfacial undercoolings compared with the SCS, as shown in the inset of Fig. 2. We further designed different thermal gradients to test the PCS in an identical system so as to confirm that the coincidence of interface positions indicates the coincidence of interface temperatures, although the difference in thermal conductivities of alumina suspension and its supernatant may affect the local thermal gradient and the interfacial position. The measured PCS under different thermal gradients keeps the same, as shown in Fig. S4 (Supplementary Information). The theoretical prediction of PCS is approximately $10^{-6}$ K (prediction A in Fig. 2 and Table 2), which is still undetectable in the present setup, but, consistent with our experimental data. Only in the extreme case, e.g., the particles of d = 1 nm and ϕ ≈ $\phi_p$ were used, the PCS could be comparable to SCS. However, in most cases, the PCS's contribution to the interfacial undercooling is minor compared with the SCS from the solvent in the solidification of colloidal suspensions.

In the above measurements, the static interfacial undercoolings were clarified. The dynamic interfacial undercooling during the freezing of colloidal suspensions, another important aspect related to the pattern formation, has never been reported before. Our experimental apparatus can also be used to quantitatively identify the dynamic interfacial undercooling[37]. Here, we measured the dynamic interfacial undercooling in the alumina suspensions of d = 50 nm, $\phi_0$ = 3.63% to further reveal the contribution of SCS and PCS. The comparison of the colloidal suspension to its supernatant can reveal the dynamic PCS, and the comparison of deionized water to the supernatant can reveal the dynamic SCS.

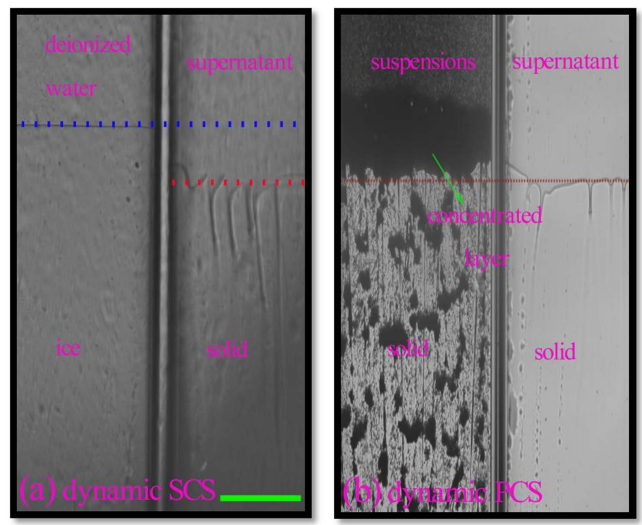

**Figure 3.** The steady-state interfacial positions in two side-by-side Hele-Shaw cells of the deionized water and the supernatant from alumina suspensions of d = 50 nm, $\phi_0 = 3.63\%$ (**a**); and two side-by-side Hele-Shaw cells of the alumina suspensions and its supernatant (**b**) in a uniform thermal gradient of G = 7.23 K/cm. The distances between the steady-state interfacial positions reveal the dynamic interfacial undercoolings. The pulling speed is V = 8.217 μm/s. The scale bar is 200 μm.

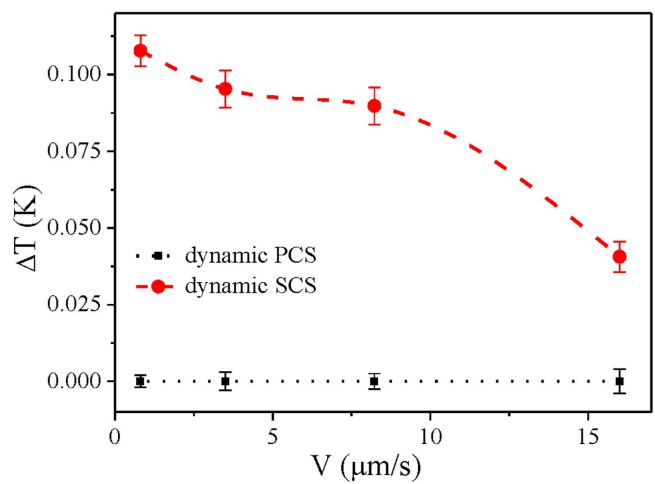

**Figure 4.** Measured dynamic SCS and PCS for alumina suspensions with d = 50 nm, $\phi_0 = 3.63\%$ under different pulling speeds; G = 7.23 K/cm.

Figure 3 shows the steady dynamic interface positions of the supernatant, deionized water and colloidal suspension under V = 8.217 μm/s and G = 7.23 K/cm. The steady state of dynamic PCS was verified as shown in Movie S1 (Supplementary Information). The comparison between the interfacial positions of the supernatant and deionized water indicates a dynamic SCS of 0.09K, as shown in Fig. 3(a). However, Fig. 3(b) shows that the dynamic PCS is undetectable, although the particles have accumulated in front of the advancing freezing interface and formed an obvious concentrated layer. Similar to the static case, the dynamic PCS is also minor compared with the obvious dynamic SCS. Therefore, the concentrated particle layer seems unable to cause an obvious PCS.

Even with different pulling speeds, the results regarding dynamic PCS are similar to that shown in Fig. 3 (shown in Supplementary Information, Fig. S5). The concentrated layer of particles in front of the freezing interface scarcely causes dynamic PCS under different pulling speeds. In contrast, the dynamic SCS varies with the pulling speed. To reveal the dynamic SCS, the interfacial position comparisons are shown in Fig. S6 (Supplementary Information). Fig. 4 shows the variation in the dynamic SCS with different pulling speeds. The increase in the dynamic SCS with decreased pulling speed is consistent with the classical alloy solidification principle[47]. This result indicates that, in the dynamic case, the SCS still plays a dominant role compared with the minor dynamic PCS for cases of different pulling speeds.

Based on the above systematic measurements, the PCS is minor in both static and dynamic cases; in contrast, the effect of SCS caused by additives in the solvent is dominant. Accordingly, the thermodynamic effect of particles from the PCS is not the fundamental physical mechanism for cellular growth and lamellar structure that are associated with the solidification of colloidal suspensions[48] which inevitably contains a large number of solutes, especially in the ice-templating method. Nevertheless, when the effect of solutes is absent, some other effects such as the force interactions between particles and freezing interface perhaps should be considered to reveal the pattern of intermittent lenses[48]. The present experimental results clearly demonstrate that the effects of additives are dominant in the ice-templating process[12,13,17,48].

## Conclusions

We considered the puzzling phenomenon of interfacial undercoolings in the solidification of colloidal suspensions *via* quantitative measurements of solute constitutional supercooling (SCS) and particulate constitutional supercooling (PCS). Based on systematic quantitative experimental measurements of both static and dynamic cases within different systems of colloidal suspensions, we found that the interfacial undercooling mainly comes from SCS caused by the additives in the solvent, while the PCS is minor. The results imply that the thermodynamic effect of particles from the PCS is not the fundamental physical mechanism for pattern formation of cellular growth and lamellar structure that are associated with the solidification of colloidal suspensions. These fundamental findings can greatly enhance our understanding of the physics of freezing colloidal suspensions in the ice-templating method, a general method to produce novel and advanced biomaterials as well as multifunctional-materials, and pave the way toward controlling pattern formation in freezing colloidal suspensions[48].

## Acknowledgements

We are very grateful for the helpful suggestions and discussions with Prof. M G Worster, Prof. S Deville and Dr. S.S.L. Peppin as well as the anonymous reviewers. This research has been supported by Nature Science Foundation of China (Grant Nos 51371151 and 51571165), Free Research Fund of State Key Laboratory of Solidification Processing (100-QP-2014), the Fund of State Key Laboratory of Solidification Processing in NWPU (13-BZ-2014) and the Fundamental Research Funds for the Central Universities (3102015ZY020).

## Author Contributions

L.W. and Z.W. designed the research. J.Y. performed the experiments. J.W., J.L. and X.L. processed experimental data. J.Y. and Z.W. wrote the paper. W.H. revised the paper. All authors analyzed and discussed the results.

## Additional Information

**Supplementary information** accompanies this paper at http://www.nature.com/srep

**Competing financial interests:** The authors declare no competing financial interests.

**How to cite this article**: You, J. *et al.* Interfacial undercooling in solidification of colloidal suspensions: analyses with quantitative measurements. *Sci. Rep.* **6,** 28434; doi: 10.1038/srep28434 (2016).